\documentstyle[12pt,psfig]{article}
\def\lsim{\raise0.3ex\hbox{$<$\kern-0.75em\raise-1.1ex\hbox{$\sim$}}}
\def\gsim{\raise0.3ex\hbox{$>$\kern-0.75em\raise-1.1ex\hbox{$\sim$}}}

\def\pom{{I\!\!P}}

\def\beq{\begin{equation}}
\def\eeq{\end{equation}}
\def\bea{\begin{eqnarray}}
\def\eea{\end{eqnarray}}
\def\bq{\begin{quote}}
\def\eq{\end{quote}}

\def\gappeq{\mathrel{\rlap {\raise.5ex\hbox{$>$}}
{\lower.5ex\hbox{$\sim$}}}}

\def\lappeq{\mathrel{\rlap{\raise.5ex\hbox{$<$}}
{\lower.5ex\hbox{$\sim$}}}}

\def\Toprel#1\over#2{\mathrel{\mathop{#2}\limits^{#1}}}

\begin{document}
\pagestyle{empty}
\begin{flushright}
{CERN-TH/2001-265}\\
\end{flushright}
\vspace*{5mm}
\begin{center}
{\bf ASPECTS OF THE UNITARIZED SOFT MULTIPOMERON APPROACH IN DIS AND
DIFFRACTION.} \\
\vspace*{1cm}
M.B. Gay Ducati$^{(a)}$, { E.G. Ferreiro$^{(b)}$,
M.V.T. Machado$^{(a)}$, C.A. Salgado$^{(c)}$}\\
\vspace{0.3cm}
{$^{(a)}$ \rm Instituto de F\'{\i}sica, Universidade Federal do Rio
Grande do Sul,\\ Caixa Postal 15051, CEP 91501-970, Porto Alegre, RS, Brazil,}\\
$^{(b)}$ \rm Depto. de F\'{\i}sica de Part\'{\i}culas,
Universidade de Santiago de Compostela \\
E-15706 Santiago de Compostela, Spain,\\
{$^{(c)}$ \rm CERN CH-1211 Geneva 23, Switzerland.}
\vspace*{2cm}\\
{\bf ABSTRACT} \\ \end{center}
\vspace*{5mm}
\noindent
We study in detail the main features of the unitarized Regge model (CFKS),
recently proposed to describe the  small-$Q^2$ domain. It takes into account
a two-component description
with two types of unitarized contributions:
one is the multiple Pomeron
exchanges contribution, interacting with the large dipole size
configurations, and the other one consists on a unitarized dipole cross
section, describing the
interaction with the small size dipoles.
Its extrapolation to higher
virtualities is performed, analyzing the ratio between soft and hard pieces
and comparing the resulting dipole cross section to that
from the saturation model. Diffraction dissociation is also considered,
showing the scaling violations in diffractive DIS and estimating the
corresponding logarithmic slope.

\vspace*{2cm}
\noindent
\rule[.1in]{16.5cm}{.002in}

\noindent
\vspace*{0.5cm}

\begin{flushleft} CERN-TH/2001-265 \\
October 2001.
\end{flushleft}
\vfill\eject

\setcounter{page}{1}
\pagestyle{plain}

\section{Introduction}

The study of a new regime of QCD, that of high density of partons, has
drawn much attention in the 
last
years. The key discovery was the
observation at HERA of the fast growth of
parton densities (mainly gluons) as the energy increases in experiments of
deep inelastic scattering.
Taking $\sigma^{tot}\sim s^{\alpha(0)-1}$ ($F_2\sim x^{-\alpha(0)+1}$),
values of $\Delta\equiv \alpha(0)-1$ in the range $0.1$ -- $0.5$ have been
reported, depending on the virtuality $Q^2$ of the photon.
However, this steep growth should be tamed, leading to the expected
limit given by the  Froissart bound ($\sigma\ \lsim\  (\log s)^2$ as
$s\to\infty$)\cite{frois}.
This boundary has been derived from very general
properties of the S-matrix, namely unitarity. A cross section growing as any
positive  power of $s$ would violate unitarity at asymptotic energies.
Thus, theoretically, some kind of saturation of this growing due to unitarity
effects 
is
expected \cite{Mueller90}. The dynamics of such very
dense partonic systems is very interesting and has been studied by many authors
both
in DIS \cite{satmodels}
and in high energy nuclear interactions \cite{satmodels2}.

The description of the $\gamma^*p$ collision in the frame where the proton
is at rest is very  appropriate to include unitarity corrections.
In this frame, the virtual photon
$\gamma^*$ emitted by the incoming lepton fluctuates into a
$q\bar q$ pair. This system then suffers multiple
interactions with the proton. Such multiple interactions restore
unitarity even in the case where it would be violated in a single collision.
In the model developed in \cite{CFKS1,CFKS2}, all these corrections
have been taken into account, and their strength is constrained by
diffractive data. Therefore, the ratio $\sigma^{diff}/\sigma^{tot}$
is related to unitarity corrections. This is a common feature to any
realization of the Gribov model \cite{Gribov}, where the amount of
rescatterings is related to diffractive production by means of AGK-cutting
rules \cite{AGK}.

In parton language, the increasing number of gluons in a proton as
$x\to 0$ makes gluon fusion very probable.  This fusion produces gluons of
higher longitudinal momentum, stopping the growing of those with the smallest
$x$. In this way unitarity is not
violated.  Such a procedure was implemented on theoretical grounds
from QCD through the multiladder exchange 
using
the GLR formalism
\cite{satmodels}, giving rise to non-linear effects in the standard linear
DGLAP approach. The outstanding quantity emerging from the unitarization
procedure is the saturation scale $Q_s^2(A,x,b)$, setting the region where
saturation phenomenon starts to be meaningful.  The
QCD-inspired  phenomenological model \cite{GW}, for instance, introduces a
quite clear identification for this scale $Q^2_s(x) \sim 1/R_0^2(x)$. There
the saturation radius $R_0(x)$, related with the mean transverse distance
between partons,  is properly extracted from the small-$x$ data from HERA.

In any of the descriptions, the unitarity corrections are
given by the non-linear terms, and  a phenomenon of saturation is expected when
these terms become important.  Since the gluons are the partons
driving  of the high energy processes, the signals of the saturation effects should
appear in the observables probing the gluonic content of the proton (or the
nucleus) \cite{AGDGPRD}. In the nuclear case, the gluon density is $\sim
A^{1/3}$ higher than in the proton. This makes unitarity corrections more
important for nuclei, producing the well-known shadowing of $F_2$
\cite{GDGPLB}. Saturation will thus start at smaller energies in nuclei
than in protons. Such a fact is the main reason
for the increasing interest in the forthcoming $eA$ experiments, where the
nucleus will be studied at energies higher than currently available \cite{GDGPLB}.

The open question is if the unitarity corrections have already shown up at
present energies and if the saturation has been reached. In
particular, at HERA, they should appear in the small-$x$ and small-$Q^2$
data \cite{AGDLPLB}. There are several proposals in this direction \cite{GW}
\cite{saturphen},
mainly for the case of
heavy-ion collisions \cite{saturnucl},
but a definitive answer is still missing. The
main difficulty that we are faced with
is the saturation scale $Q_s$, staying in the
transition interval of 1--2 GeV, which leads the effects to be hidden in more
inclusive observables. In this kinematical region the standard QCD perturbative
expansion is expected not to be completely reliable. For instance, higher
twist terms to the linear approach should be taken into account in such a domain.
Moreover, this 
is the transition region
between the soft and hard
domains, i.e. the perturbative approaches (including saturation or properly
adjusting initial conditions) and the Regge-inspired models are competing, and
both frameworks seem to describe the current small-$x$ data.

Bearing in mind that the saturation phenomenon is required in a complete
understanding of the high energy reactions, and that a consistent treatment of
both inclusive and diffractive processes should be taken into account,  in
this work we study derivative quantities using the Regge unitarized  CFKS model
\cite{CFKS1,CFKS2}. In this hybrid model, both soft (multiperipheral
Pomeron and reggeon exchanges) and hard (dipole picture) contributions are
properly unitarized in an eikonal way. This approach describes
the transition region and can be used as initial condition for a QCD
evolution at high virtualities \cite{novo}.
The extrapolation to the higher-$Q^2$ domain
is also  performed here, 
checking the behaviour of the model without including QCD evolution.
We discuss the similarities and/or connections  with
the phenomenological saturation model \cite{GW}, stressing that a QCD
evolution is required for a correct description of higher $Q^2$ in the
inclusive case. For the diffractive case, such a procedure is not formally
required, since the non-perturbative sector is dominant in this case. The
diffractive structure function is extrapolated to the available larger-$Q^2$
range. In particular, the diffractive logarithmic slope, which has been claimed
as a possible new observable to disentangle dynamics \cite{slope1,slope2}, 
is
calculated and compared with the result from the saturation model.

\section{The inclusive case}

We start by briefly reviewing the CFKS approach. It  interpolates between
low
and high virtualities $Q^2$, which are related to the dipole separation size,
$r$,  at the target rest frame,  considering a two-component
model \cite{CFKS1,CFKS2}. Considering the unifying picture of the
color dipoles, the separation into a large size (in \cite{CFKS2} it is
called $L$) and a small size (called $S$ in
\cite{CFKS2}) components of the $q\bar{q}$ pair is made in terms of the
transverse distance $r$ between $q$ and $\bar{q}$. The border value,
$r_0$, is treated as a free parameter - which turn out to be $r_0 \sim
0.2$~fm. This value agrees with the correlation length of nonperturbative
interactions observed in lattice calculations.
Hereafter we use the notation
{\it soft} for the large size configuration and {\it hard} for the small size one.
This separation corresponds, in Ref. \cite{CFKS1}, to the separation of the 
$q\bar{q}$ pair fluctuation into the ``aligned'' component 
(large transverse size, soft component)
with a strongly
asymmetric
sharing of the momentum fraction between $q$ and $\bar{q}$, and the
``symmetric'' component (small transverse size, hard component).

The soft component considers multiple 
Pomeron exchanges (and
reggeon $f$) implemented in a quasi-eikonal approach \cite{martirosian}. 
It also includes the resummation of triple Pomeron branchings (the so-called
fan diagrams).
The initial input is a
phenomenological Pomeron with fixed intercept 
$\alpha_P(0) = 1 + \varepsilon_{\pom}=1.2$
(further changes are due to absorptive corrections), and an
exponential parametrization for the $t$ dependence is considered. In the
impact parameter representation,  the $b$-space,  it looks like (in
photoproduction $Q^2=0$):  \begin{eqnarray} \chi^{\pom}(s,b) \simeq C_{\pom}
\frac{f_{\pom}}{B_{el}(s)}\, \left(\frac{s}{s_0}\right)^{\varepsilon_{\pom}}\,
\exp[-b^2/B_{el}(s)]\,, \end{eqnarray}
where $B_{el}(s)$ is the elastic slope, which is  parametrized as in the
hadronic reactions. The $f_{\pom}$ is an effective Pomeron--proton coupling.
In the electroproduction case, the  initial input  is described in an
analogous way: \begin{eqnarray}
\chi^{\pom}(s,b,Q^2) \simeq \frac{C_{\pom}}{R(x,Q^2)}
\left(\frac{Q^2}{s_0 + Q^2}\right)^{\varepsilon_{\pom}}\,
x^{-\varepsilon_{\pom}} \exp [-b^2/R(x,Q^2)] \,, \end{eqnarray}
corresponding to the Regge parametrization for the  amplitude
of the soft Pomeron exchange,  similar to  the
Donnachie--Landshoff one \cite{DL}.
The function $R(x,Q^2)$ comes from the exponential
assumption about the $t$ dependence and further transformation to the impact
parameter representation. 
We shall remark here that, in the CFKS approach, the authors consider a Pomeron
fixed
intercept of $\alpha_P(0) = 1 +
\varepsilon_{\pom}=1.2$, $\varepsilon_{\pom}=0.2$ 
(a semi-hard value rather than a soft
one). The unitarization effects, described by multi-Pomeron exchanges, lead
to an effective intercept 
$\varepsilon_{eff} = {
d \ell n
F_2(x,Q^2) \over d \ell n \left ( {1 \over x}\right )}$ that decreases as $Q^
2$ or $x$
decreases due to the increase of shadowing effects.

The resummation of the triple-Pomeron branches is encoded in the
denominator of the amplitude $\chi^{n\,\pom}$, i.e. the Born term in the
eikonal expansion. Moreover, the corrected amplitude is eikonalized in the
total cross section,
\begin{eqnarray}
\chi^{n\,\pom}(x,Q^2,b)& = &
\frac{\chi^{\pom}(x,Q^2,b)}{1+a\chi_3(x,Q^2,b)}\,,\\
\sigma^{n\,\pom}(x,Q^2,b) & \simeq & 1-\exp \left[
-
\chi^{n\,\pom}(x,Q^2,b)\,\right]\,. \end{eqnarray}
where the constant $a$ depends on the proton-Pomeron and the triple-Pomeron
couplings at zero momentum transfer ($t=0$). Refs.
\cite{CFKS1,CFKS2} give a more detailed discussion.

The eikonalization
procedure modifies the growth of the total cross section from a steep
power-like behavior to a milder logarithmic increase. The above parametrization
corresponds to the interaction with the large size dipole configurations
and therefore dominates in low-$Q^2$ values. The total soft contribution is
 obtained by integrating over the impact parameter the cross section at
fixed $b$, $\sigma^{n\,\pom}(x,Q^2,b)$,
\begin{eqnarray}
\sigma^{soft}(s,Q^2)=4\,\int d^2b\,\,\sigma^{soft}(s,Q^2,b)\,.
\end{eqnarray}

The hard component is  considered in  the color dipole picture of DIS
\cite{dipole}. The dipole cross section, modeling the interaction
between the $q\bar{q}$ pair and the proton, $\sigma^{dipole}(x,r)$,   is taken
from the eikonalization of the  expression above $\chi^{n\,\pom}(s,b,Q^2)$
already corrected  by triple-Pomeron branching  (the fan
diagrams contributions). 
The configurations considered are those 
with a small transverse distance between the quark--antiquark pair  in the dipole.
The corresponding cross section  is extracted by considering the
contributions coming from  distances between 0 and  $r_0=0.2$ fm (1
GeV$^{-1}$) -- see discussion above --, 
whereas for $r>r_0$ the contributions are described by the
soft piece already discussed. In such small distances, perturbative QCD is
expected to work.  The total cross
section considering this dipole cross section is expressed as \cite{CFKS2}:
\begin{eqnarray} \sigma^{hard}_{tot}(x,Q^2)
& = & \int_0^{r_0} d^2 r \, \int_0^1 d \alpha \, |\Psi^{T,L}_{\gamma^*
q}(\alpha,r)|^2  \,   \sigma^{dipole}_{CFKS}(x,r) \,,  \\
\sigma^{dipole}_{CFKS}(x,r) & = & 4\,\int d^2b \,\, \sigma^{n\,\pom}(x,Q^2,b,r)\,,\\
\sigma^{n\,\pom}(x,Q^2,b,r) & \simeq & 1-\exp [
-
\,r^2 \chi^{n\,\pom}(x,Q^2,b)\,]\,,
\end{eqnarray}
\noindent
where $T$ and $L$ correspond to transverse and longitudinal polarizations
of a
virtual photon, $\Psi^{T,L}_{\gamma^*
q}(\alpha,r)$ are the corresponding wave functions of the
$q\bar{q}$-pair.

The $r^2$ dependence is introduced in the Born term of the eikonal
expansion, presented  in the last expression above, in order to ensure the
correct behavior determined by
the color transparency: for small $r$ the growth in radius should
be  proportional to $r^2$, $\sigma^{n\,\pom}(x,Q^2,b,r)  \simeq  r^2\,
f(x,Q^2,b)$. This condition, valid for fixed $s$ and $Q^2$ as $r
\to 0$, is a property of the single Pomeron exchange. Thus a factor $r^2$ has
been
introduced in the eikonal of eq. (8). 

Another difference between the soft and hard component is the fact that 
the contribution of the $f$-exchange (reggeon exchanges) to the hard component
is very small, and has been neglected.

The weight of each contribution (soft and hard) in the total cross section
[and $F_2(x,Q^2)$] can be obtained, providing an analysis of the
role played by each piece of the model. Such a procedure allows us to
explicit the regions of $x$ and $Q^2$ where the sectors contribute. In Figs.
1 and 2 we calculate the ratio $R_{SOFT}$, defining the fraction of the total
contribution arising from the soft sector:

\begin{eqnarray}
R_{SOFT}(x,Q^2)=
\frac{
\sigma_{tot}^{soft}(x,Q^2)}{
[\sigma_{tot}^{soft}(x,Q^2)+\sigma_{tot}^{hard}(x,Q^2)]}\,. \end{eqnarray}

\begin{figure}[t]
\centerline{\psfig{file=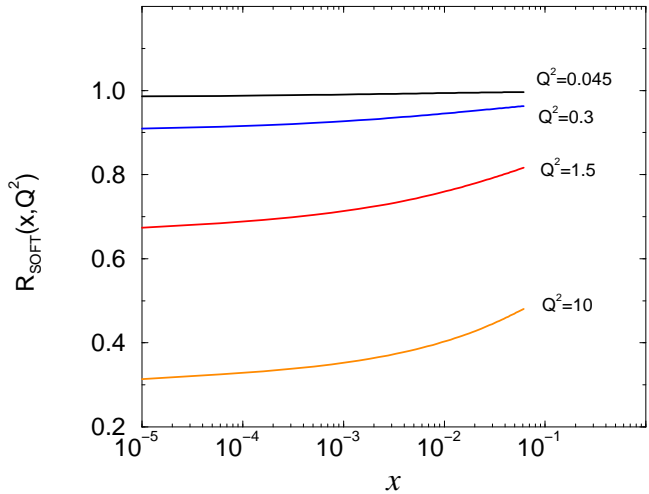,width=90mm}}
\caption{The ratio $R_{SOFT}$ as a function of $x$ at fixed virtualities.}
\end{figure}

$ $ From Fig. 1 we note that the soft contribution slowly increases as
the momentum fraction $x$ goes to higher values, almost independently of the
virtuality $Q^2$. This is due to the fact that higher reggeon trajectories
$f$ are included in the soft part, but not in the hard one. Regarding fixed
virtualities, the soft piece dominates completely the total cross section at
$Q^2=0.045$. As $Q^2$ increases the contribution goes down. For instance, at
$Q^2=10$ GeV$^2$ it contributes about half of the cross section. Extrapolating
up to higher virtualities, the soft piece saturates at about  5--15\% of the
total result.

\begin{figure}[t]
\centerline{\psfig{file=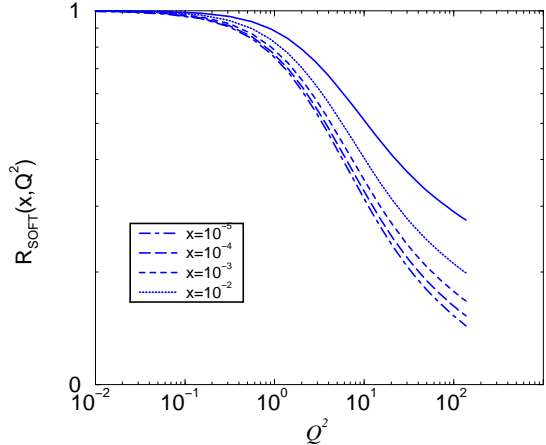,width=80mm}}
\caption{The ratio $R_{SOFT}$ as a function of $Q^2$ at fixed momentum
fraction $x$.}
\end{figure}

 $ $ Figure 2 clearly shows that the soft piece is  dominant
at  $Q^2=0.01$ and decreases as the virtuality grows. The behavior is
monotonic, almost independent of the momentum fraction $x$. For instance, at
$Q^2=100$ GeV$^2$, it contributes with 20\% at $x=10^{-2}$ and 5\% at
$x=10^{-5}$. Such a reduction on the soft content is related to the  coupling
of the photon to the asymmetric dipoles $g^2_{soft}(Q^2)\sim 1/ (1+ Q^2/
m_{soft}^2)$ and to the enhancement in $Q^2$ provided by the photon
wave function (at high $Q^2\gg Q^2_s(x)$ the symmetric dipole configuration
provides the scaling with logarithmic violation).

An interesting issue  is the relation between the dipole cross section
coming from the CFKS model and the phenomenological one of
G.-Biernat-W\"{u}sthoff \cite{GW}. The GBW cross section is parametrized as:
\begin{eqnarray}
\sigma^{GBW} (x,r) & = & \sigma_0 \left[ 1- \exp(-r^2/4R^2_0(x))\right]\,,\\
R^2_0(x) & = & \left(\frac{x}{x_0}\right)^\lambda \rm{GeV}^{-2}\,,
\end{eqnarray}
where $\sigma_0=23.03$ mb properly normalizes the dipole cross section.  The
remaining parameters are $\lambda=0.288$ and $x_0=3.04\times 10^{-4}$,
all of them determined from the small-$x$ HERA data.  The $R_0(x)$ is
the main theoretical contribution, defining the saturation scale, which is
related with the taming of the gluon distribution at small $x$ (unitarity
effects) \cite{satmodels}. The above expression has been used to describe both
inclusive and diffractive structure functions, in good agreement with the
experimental results. The comparison between this approach and the
CFKS dipole cross section is shown in Fig. 3.  We have  plotted the
adimensional result, since the normalization for the CFKS dipole cross
section, $\sigma_0$, is not determined from data. Indeed, for a comparison
with experiment using only the hard piece from CFKS, the adjustable parameters
would have to
be refitted. We consider here that this can be absorbed by a suitable
normalization, and carry the $r$ interval beyond the range set by the model
($r<r_0$).  The main feature of the GBW parametrization is that it ensures that
the
dipole cross section  grows linearly with $r^2$ at small transverse separation,
whereas it saturates at large size configurations. The picture emerging from
the CFKS is slightly different,  presenting a mild (logarithmic) increase with
$r$, away from huge separation sizes that shifts the saturation scale up to
very high virtualities. Although the continuous and smooth increasing with
the radius, in the CFKS approach the cross section  underestimates the GBW one
for all $r$.

A comment on the normalization is in order.
The GBW formula would correspond to the hard part of CFKS without
triple-pomeron ($a=0$), and taking a step function for the profile instead of a
gaussian. This makes unitarity corrections stronger. 
In any case, we can compare GBW and the hard part of CFKS: taking $a=0$
and $\exp[-b^2/R(x,Q^2)]\longrightarrow \Theta[b^2-R(x,Q^2)]$ in CFKS,
doing the integral in $b$ and comparing with GBW,
one obtains $\sigma^{dipole}_{CFKS}(x,r)=
\sigma_0\, \left[ 1- \exp(-r^2\, \chi^{\pom}(s,b,Q^2)\right]$, with
$\sigma_0=\pi R(x,Q^2)\sim $ 20 mb, in agreement with GBW value.
This
value, however, depends logarithmically on $x$ and $Q^2$, because of the
increase of the proton radius, which is taken into account in (2). The
comparison between the exponents (the eikonals, which contain most of the
parameters in CFKS case)
of the above expression
and of the eq. (10)
is less clear, as the $x-$dependences are
different and the triple pomeron can not be neglected in this case.

\begin{figure}[t]
\centerline{\psfig{file=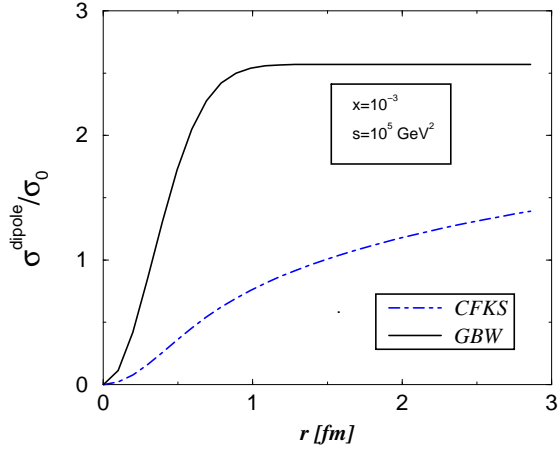,width=80mm}}
\caption{ The comparison between the saturation dipole cross section from
G.-Biernat-Wusthoff (GBW) and CFKS as a function of the transverse dipole
separation $r$ at fixed $x$ ($s$).}
\end{figure}

\section{The diffractive case}

 The diffractive sector in the CFKS approach is constructed by a
three-component
model \cite{CFKS1,CFKS2}, using the AGK cutting rules to relate the
elastic multiple scattering amplitude to the inelastic diffractive
contribution \cite{AGK}. The first term comes directly from the soft piece, the
second one from the triple-Pomeron (and the reggeon $f$) interaction and the
last one from the hard (dipole) piece. We notice that these contributions
define only the energy, $s$ (and momentum fraction $x$), and the virtuality
dependences. The spectrum on $\beta$ is introduced by hand, based on earlier
soft and hard (pQCD) calculations. The first component is written as:

\begin{eqnarray}
F^D_{2\,(soft)}(x,Q^2, \beta)\sim F^{D\,(\rm{Born})}_{soft} \, K_L(s,Q^2)\,
\beta^{-\varepsilon_{\pom}} \, (1-\beta)^{n_p(Q^2)}\,,
\end{eqnarray}
 where $F^D{\,(\rm{Born})}_{soft}\sim \chi^{n\,i}(s,Q^2)  \chi^{n\,k}(s,Q^2)$ is the
lowest-order (Born) approximation for that function, with $i,k=\pom, f$.  The
suppression  factor due to higher order multipomeron exchanges is
$K_L(s,Q^2)= \sigma^{(0)}_{soft}/ \sigma^{(0)\,\rm{Born}}_{soft}$, with
$\sigma^{(0)}_{soft}=4g^2_L(Q^2)\int d^2b \, [\sigma_{soft}(s,Q^2,b)]^2$.
Further details can be found in \cite{CFKS2}.  The $\beta$ dependence is taken
from the typical CKMT Pomeron structure function, which is connected with the
deuteron structure function by the identification $x \rightarrow \beta$
\cite{CKMT}.

The hard contribution is expressed as:
\begin{eqnarray}
F^D_{2\,(hard)}(x,Q^2, \beta)\sim \sigma^{(0)\,L}_{hard}\,
\beta^3(1-2\beta)^2 + \sigma^{(0)\,T}_{hard} \, \beta^2(1-\beta) \,,
\end{eqnarray}
where the $\beta$ dependence comes from a pQCD guess for  the Pomeron
structure function \cite{BFS}. Also, $\sigma^{(0)\,T,L}_{hard}=\int d^2b \,
[\sigma^{T,L}_{hard}(s,Q^2,b)]^2$. However, the $\beta$ spectrum is slightly
different from the  most recent pQCD
calculations, where the transverse contribution behaves like $\sim
\beta(1-\beta)$ \cite{BW}. The hard component dominates at large $\beta$ in
the CFKS approach, mainly concerning the transverse component, where the
charge in the $\beta$ dependence will no significantly modify the data
description, moreover due to few data to constraint the adjustable parameters
in this region.

Regarding the $\beta$ dependence, the region for medium values ($\beta\sim
0.4$)  is dominated by the soft term, which in pQCD is associated to the
transverse photon contribution \cite{BW}. The small $\beta$ region is dominated
by the triple-Pomeron piece, in agreement with the pQCD expectations, which is
obtained by considering the higher twist $q\bar{q}$+gluon configuration.
Moreover, the hard contribution is leading in the large-$\beta$ region,
associated in this case with a suppression of the transverse contribution and
an  enhancement of the longitudinal piece in comparison with the expected pQCD
behavior \cite{BW}.

\begin{figure}[t]
\centerline{\psfig{file=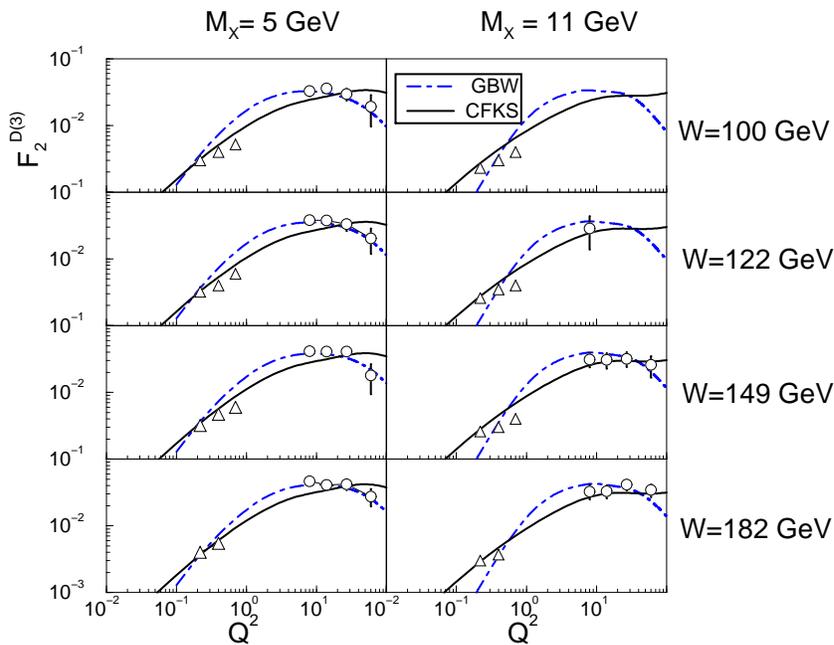,width=100mm}}
\caption{ The diffractive structure function as a function of $Q^2$ at fixed
$M_X$ and $W$. The preliminary data are from ZEUS Collaboration (triangles).
The published data are the circles. The CFKS and GBW results are shown in
the same plot.} \label{f2dzeus} \end{figure}

The CFKS approach describes with good agreement the diffractive DIS data in
the broad range $0<Q^2<18$ GeV$^2$. In order to study the model in comparison
with the pQCD approaches, here we extrapolate the prediction for the
diffractive structure function of the CFKS approach for higher values of the
virtuality.
We use the preliminary ZEUS analyses, considering the $Q^2$ dependence
at fixed mass $M_X$ and centre-of-mass energy $W$ \cite{prelZEUS}. These data
provide information at both small and large virtualities bins. 
It is interesting to compare the predictions of the CFKS model and the
saturation model
\cite{GW}
for the diffractive structure function. Both models are depicted in
the plots of Fig. \ref{f2dzeus}.

The agreement of CFKS approach with data is remarkable
even at high virtualities,
where the model is expected not to be reliable. However the interpretations at
low $Q^2$ are quite different. In the saturation model, the reliability of the
pQCD calculation is extended to smaller virtualities through the saturation
scale $R_0(x)$, where the dependence is mostly due to the longitudinal
photon configuration, by the higher twist $q\bar{q}$+gluon. Instead, in
the CFKS model the main contribution in the region of interest comes from the
soft triple-Pomeron contribution.

As a final study, we perform the calculation of the $Q^2$ logarithmic slope of
the diffractive structure function $F_2^{D(3)}$.
 The motivation is that this
observable is a potential quantity to distinguish soft and hard dynamics
in diffractive DIS \cite{slope1,slope2}. The measurements of the derivative
quantity $F_2$-slope on $Q^2$  have allowed a renewed interest on
testing the matching of hard and soft approaches and have
provided constraints for
the saturation formalisms. The reported turnover on the $x$ dependence
has been
associated with the transition region between the soft and hard domains.
 When we focus in diffractive DIS, in particular
the structure function $F_2^D$, the situation is far from clear: initially
considered as a predominantly soft process, the experimental results suggest
that the diffractive cross section at HERA contains hard and soft
components.

Still,
the diffraction stands a more profitable field to study saturation effects than
the inclusive case.
This comes from the fact that, in DDIS,
the large dipole size configuration (soft content) is more relevant
than in the DIS
reaction\cite{GW}.
Although
there are quite different approaches, based on different physical dynamics,
 applied to the interpretation of  the diffractive measurements,
almost all of them fit the data set
properly\cite{H1diffnew}. Therefore
a derivative
quantity, the diffractive logarithmic slope, has been proposed.
It can help to distinguish
the underlying dynamics in diffractive DIS, settling the validity range
of the different approaches, if
such observable is measured. Here, we have calculated
the slope as a function of the Pomeron momentum fraction $x_{\pom}$ and
we have performed
a comparison between the CFKS  and the G.-Biernat-Wusthoff (GBW)approaches.

In this calculation, the main feature of the GBW approach is the
presence of positive and negative slopes,
contrary to the pQCD non-saturated case \cite{slope2}.
In the pQCD model without saturation, the parameters fixed by the
previously available
data lead
to a predominantly positive slope for all kinematical region, even at large
$Q^2$ (large $\beta$). We notice that this situation can be changed by a
further analysis, considering the updated measurements of H1
\cite{H1diffnew}, which have enlarged the available kinematical range and
have provided new measurements for the regions previously covered.

It is important to emphasize that a transition in the slope
takes place in the preliminary ZEUS analyses of diffractive DIS
\cite{prelZEUS} at large virtualities, where the saturation model \cite{GW} is
considered to describe the $Q^2$ dependence of the diffractive structure
function.
In this model, the analysis is performed using
$M_X$ and $W$ as kinematical variables instead of $\beta$ and $x_{\pom}$,
due to the similarity
of the behaviors of $d\sigma/dM_X$ and $ \sigma_{tot}(\gamma^*p)$ in the same
kinematical range.
There, the growth of $x_{\pom}F_2^D$ versus $Q^2$
is stopped at $Q^2\sim 10$ GeV$^2$ and decreases smoothly for larger
virtualities. The transition region corresponds to $\beta \sim 0.2$ for
$M_X=5$ GeV and $\beta \sim 0.07$ for $M_X=11$ GeV \cite{slope2},
$\beta=Q^2/(Q^2+M_X^2)$. These features can be verified observing 
the plots of Fig.
(4) at large virtualities.

\begin{figure}[t]
\centerline{\psfig{file=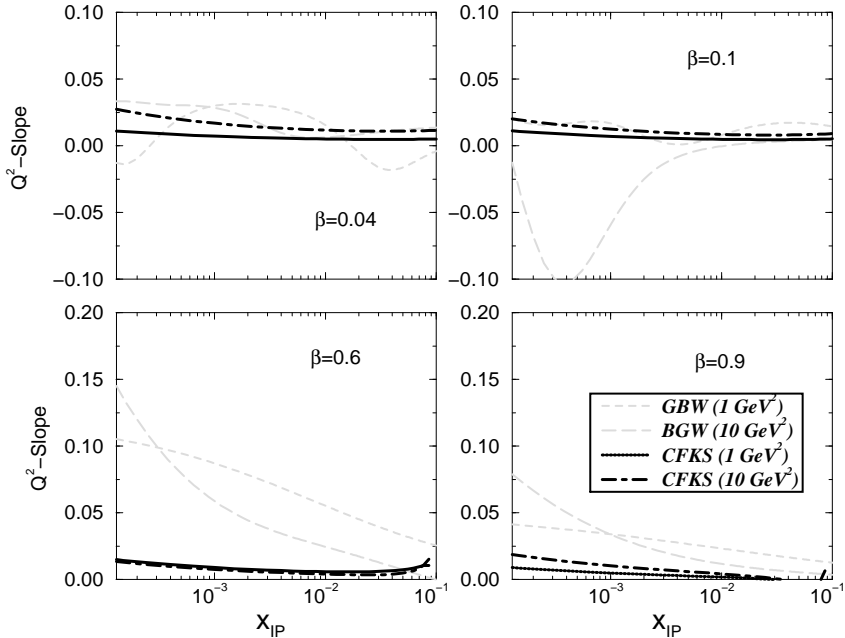,width=100mm}}
\caption{ The $Q^2$ logarithmic slope of the diffractive structure function as
a function of $x_{\pom}$ at fixed $\beta$, for
$Q^2=1$ and $Q^2=10$ GeV$^2$.
The CFKS and GBW results are shown in the same plot.}
\label{slopeq2plot}
\end{figure}

Here, we compared the results from the saturation model with those from
the model analyzed in this work. They are shown in Fig. (\ref{slopeq2plot}),
where the slope is calculated as a function of $x_{\pom}$ for fixed $\beta$
and at two different virtualities. We choose $Q^2=1 $ and 10 GeV$^2$ because
this is the region where the CFKS model is formally valid.
However, we emphasize
that it can be extended to higher virtualities in the diffractive case, since
the soft component is stronger than in the inclusive case.
The
saturation model produces a transition between positive and negative slope
values at low $\beta=0.04$ (upper plots), while it shows a positive slope for
medium and large $\beta$. Instead, the CFKS approach  presents a positive
slope for the whole $Q^2$ and $x_{\pom}$ ranges, flattening at large $\beta$,
similarly as the non-saturated pQCD calculations. These features can be probed
if the slope is measured and
could help the  understanding of the underlying dynamics.
Finally, the results above, mainly those ones for the $F_2^D$
structure function, corroborate the CFKS model as a consistent hybrid approach
to describe diffractive DIS, with a close connection to the above mentioned
pQCD
calculations.

\section{Conclusions}

A deeper understanding of the saturation phenomenon is required to perform
reliable estimations for the  current and forthcoming  high energy reactions.
The saturation scale, which sets the onset of the unitarity corrections, is
found to be in the transition regime of low $x$ and $Q^2$. In this
domain, both Regge-inspired phenomenology and improved pQCD calculations
(perturbative shadowing, higher twist), 
considering unitarity effects, are able
to describe the data with high precision.  
The most advantageous ones are those
describing in an unified way the inclusive processes as well as diffractive
ones. In this letter we have considered the two-component multireggeon model of
\cite{CFKS1,CFKS2} and calculated some related quantities.

The ratio of the
soft content in the model has been 
calculated, verifying that it dominates at low
$Q^2$, diminishing at higher virtualities. This shows
that the unitarity corrections in this model
are more important in the soft component
than in the hard one. Moreover, these corrections are higher twist at large
$Q^2$ in the second case.

We have also studied the robustness of the CFKS model 
to describe a large range in $Q^2$ without to consider a pQCD evolution.
A good hint to answer this question 
comes from the analysis of
the hard piece (symmetric
dipole configurations), in particular the corresponding dipole cross
section. We have found that the $r$-saturation  of this quantity lies larger at
values of the radius  than in the phenomenological GBW model. A similar
consideration is far from clear for the diffractive case, where the
non-perturbative (soft) component plays a more important role.

We have extrapolated 
the estimations for the diffractive structure function at
high virtualities, verifying
that a broad description is obtained, and that it is
in reasonable agreement with the saturation pQCD model. 
An additional quantity
has been proposed in 
order to describe the dynamics of the diffractive dissociation
\cite{slope1,slope2}, in particular 
the diffractive slope. It has been calculated using the
CFKS model, and its main feature is a similar behavior to the one  predicted
by pQCD calculations \cite{BW}.

\section*{Acknowledgements}
M.B.G.D and M.V.T.M are supported by CNPq and by PRONEX (Programa de Apoio a
N\'ucleos de Excel\^encia), Brazil. E.G.F is supported by
the contract AEN99-0589-C02 of CICYT
of Spain.
C.A.S. is supported by a Marie Curie
Fellowship of the European Community program TMR (Training and Mobility of
Researchers), under the contract number HPMF-CT-2000-01025.

\end{document}